\documentstyle[preprint,aps]{revtex}
\begin{document}

\title{Differential and partial cross sections of elastic
and inelastic positronium-helium-atom scattering} 

\author{Sadhan K. Adhikari}
\address{Instituto de F\'{\i}sica Te\'orica, 
Universidade Estadual Paulista, 
01.405-900 S\~ao Paulo, S\~ao Paulo, Brazil}

\date{\today}
\maketitle

\begin{abstract}

Scattering of 
positronium (Ps)  by helium  atom  has been investigated in a
three-Ps-state
coupled-channel model including  Ps(1s,2s,2p) states
 using a recently proposed time-reversal-symmetric regularized
electron-exchange model potential.
Specifically, we report results of differential cross sections for elastic
scattering and  target-elastic  Ps excitations. 
We also present results for  total and  different
partial
cross sections and compare them with experiment and other calculations.

{\bf PACS Number(s):  34.10.+x, 36.10.Dr}

\end{abstract}



Scattering of 
exotic 
ortho-positronium  atom with long life time (142 ns)  by neutral gas atoms
and molecules is
of
fundamental interest in both physics and chemistry. Recent high precision
measurements of positronium (Ps) scattering by H$_2$, N$_2$, He, Ne, Ar,
C$_4$H$_{10}$, and C$_5$H$_{12}$ \cite{1,1a,2,3,3x,3a}  have enhanced 
theoretical activities \cite{4,4a,5,5x,5y}
in this subject. 
Due to internal symmetry the direct static
Born potential for elastic and even-parity transitions for these processes
is zero and exchange correlation plays an important role for a correct
description at low energies \cite{5x,5y}.

Recently, we suggested \cite{6} a regularized nonlocal electron-exchange
model
 potential with a single parameter $C$ and used it in the
successful study of of Ps scattering by H \cite{7,7a}, He \cite{6,15}, Ne
\cite{15}, Ar \cite{15} and H$_2$ \cite{16,17}. Our results were in
agreement with experimental total cross section \cite{1,2}, specially at
low energies for He, Ne, Ar and H$_2$. In our initial calculations we used
a non-symmetric form of the model exchange potential for Ps scattering by 
H \cite{7a}, He \cite{6},
and H$_2$ \cite{16}. Subsequent studies yielded improved results 
 with a  time-reversal  symmetric form of the model
potential for Ps scattering by H \cite{7} and H$_2$ \cite{17}. For H it
was found
\cite{7} that 
the
symmetric potential yielded excellent results for S-wave singlet Ps-H
binding and resonance energies in agreement with accurate variational
calculations \cite{18}.  The symmetric potential also led to very good
results \cite{15} for
low-energy cross sections for Ps scattering by He, Ne, Ar, and H$_2$ in
excellent agreement with experiment \cite{2}.

The problem of Ps-He scattering is of relevance to both experimentalists
and theoreticians. Theoretically, it is the simplest of all Ps-scattering
problems, which has reliable experimental cross sections. Once a good
theoretical understanding of this system is obtained, we can try to
understand the problem of Ps scattering by complex atoms and molecules. 
With this objective we reinvestigate the problem of Ps scattering by He at
higher energies using the time-reversal symmetric form of the exchange
potential. We consider the three-Ps-state coupled-channel model with
Ps(1s,2s,2p) states for calculating different elastic and inelastic cross
sections of Ps-He scattering. We calculate the different Ps-He
differential cross sections which are of great interest to
experimentalists \cite{3x}, in addition to the different angle-integrated
partial cross sections. The differential cross sections carry detailed
information about the scattering process.  
Cross sections for higher excitations and
ionization of Ps are calculated by the Born approximation and added to the
above Ps(1s,2s,2p) cross sections to yield the total cross section which
is compared with experiment.

The theory for the coupled-channel study of Ps-He scattering
with the regularized model potential has already appeared in the
literature \cite{4,6,7,15}. It is worthwhile to quote the relevant
working
equations here. For target-elastic Ps-He scattering 
we solve the following 
Lippmann-Schwinger scattering integral
equation in momentum space for the total electronic doublet spin state 
\begin{eqnarray}
f^-_{\nu ',\nu} ( {\bf k',k})&=&
{\cal B}^- _{ \nu ',\nu}({\bf k ',k})
\nonumber
\\
&-&\sum_{\nu ''}
\int \frac{   {\makebox{d}{\bf k''}}    }   {2\pi^2}  
\frac {   {\cal B}^- 
 _ {\nu ',\nu''}  
({\bf k ',k''}) 
f^-                   _{ \nu'' ,\nu}      ({\bf k'',k}) }
{{k}^2_{\nu ''}/4-k''^2/4+ \makebox{i}0}
 \label{4}
\end{eqnarray}
where the
 Born amplitude, $B^-$,   is given by  $
 {\cal B}^-_{\nu ',\nu}({\bf k',k}) = 
 g^D_{\nu ',\nu}({\bf
k',k})- g^ E_{\nu ',\nu}({\bf k',k}),$ 
   where $g^D$  and $g^E$ represent the direct and 
exchange Born  amplitudes and  $f^ -$ 
the 
scattering amplitude, respectively. The quantum states are labeled by
the indices $\nu$ referring to 
the Ps atom. The variables ${\bf k}$, ${\bf k'}$, ${\bf k''}$ etc. denote
the appropriate momentum states; ${\bf k}_{\nu ''}$ is the on-shell
relative momentum of Ps with respect to He in  channel $\nu ''$. We
use units $\hbar = m = 1$ where $m$ is the electron mass.
  The differential cross section is given 
by 
\begin{equation}
\left(\frac{d\sigma}{d\Omega}\right)_{\nu ',\nu} = \frac
{k'}{k }
|f^ -_{\nu ',\nu}({\bf k',k})|^2.
\end{equation}

For He ground state, the space part of the Hartree-Fock (HF) wave function
is given by
$\Psi({\bf
r_1,r_2})= [\varphi({\bf r_1})\varphi({\bf r_2})]$.
 The position vectors of the electrons
are ${\bf r_1}$ and ${\bf r_2}$, and $\varphi$ is taken to be in  the form 
$\varphi({\bf r})=\sum_{\kappa} a_{\kappa } \phi_{\kappa }({\bf r})$,
where $\phi_{\kappa }({\bf r})$ are the atomic orbitals. 

The direct and exchange potentials are, respectively,  given by
\cite{6,15}
\begin{eqnarray}\label{1x} 
g^D_{\nu',\nu} ({\bf k_f,k_i})&= &
\frac{4}{Q^2}\left[
2-\sum_{\kappa, \kappa'} a_{\kappa } a_{\kappa ' 
}
\int \phi^*_{\kappa '}({\bf r})\exp ( \makebox{i} {\bf Q.
r})\phi_{\kappa  }({\bf r})
{\makebox{d}}{\bf r}  \right] 
\nonumber \\ &\times&
\int \chi^*_{\nu '}({\bf  t })\left[2\mbox{i} \sin ( {\bf Q}.{\bf  t
}/
2)\right]\chi_{\nu}({\bf  t } ) \makebox{d}{\bf  t }.
\end{eqnarray}
and 
\begin{eqnarray}\label{1} 
g^E_{\nu',\nu} ({\bf k_f,k_i})&= &
\sum_{\kappa,\kappa '}\frac{4a_{\kappa }a_{\kappa'
}(-1)^{l+l'}}{D_{\kappa\kappa '}}
\int \phi^*_{\kappa '}({\bf r })\exp ( \makebox{i} {\bf Q.
r})\phi_{\kappa } ({\bf r})
\makebox{d}{\bf r}\nonumber \\ &\times&
\int \chi^*_{\nu '}({\bf  t })\exp ( \makebox{i}{\bf Q}.{\bf  t }/
2)\chi_{\nu}({\bf  t } ) \makebox{d}{\bf  t }
\end{eqnarray}
with 
\begin{equation}\label{2}
D_{\kappa, \kappa '}=(k_i^2+k_f^2)/8+C^2[(\alpha_{\kappa
}^2+\alpha_{\kappa '
}^2)/2+(\beta_\nu^2+
\beta_{\nu'}^2)/2]
\end{equation}
where $l$  and $l'$ are  the angular momenta of the initial and final Ps
states, 
the initial and 
final Ps momenta are ${\bf k_i}$  and ${\bf k_f}$, ${\bf Q = k_i -k_f}$,
$\alpha_{\kappa }^2/2$ and $\alpha_{\kappa  ' }^2/2$, and $\beta_\nu^2$
and $\beta_{\nu '}^2$ are the binding energy parameters  of the initial
and final
  He orbital  and  Ps states in atomic units, 
respectively, and $C$ is the only parameter of the potential. Normally,
the parameter $C$ is taken to be unity which leads to reasonably good
result \cite{15,17,23}. However, it can be varied slightly from unity to
get a
precise fit
to a low-energy observable. This variation  of $C$ has no effect on the
scattering observables at high energies and the model exchange potential
reduces to the Born-Oppenheimer exchange potential \cite{19} at high
energies. 
In the present study we use
the value $C=0.84$ throughout.
  This value of $C$ leads to a very good fit of the elastic
Ps-He cross section with the experiment of Skalsey et al. \cite{2}.
This exchange potential for Ps scattering is considered
\cite{6}  to be a
generalization 
of the Ochkur-Rudge exchange potential for electron scattering \cite{20}.

After a partial-wave projection, the system of coupled equations (\ref{4}) 
is solved by the method of matrix inversion. A maximum number of partial
waves $J_{\mbox{max}}$ is included in solving the system of coupled
equations.  The differential and angle-integrated partial cross sections
so calculated are augmented by Born results for higher partial waves $J>
J_{\mbox{max}}$.  A maximum of 40 Gauss-Legendre quadrature points are
used in the discretization of
each momentum-space integral.  The calculations are performed with the
exact Ps wave functions and the HF orbitals for He ground state \cite{21}.
Although it is relatively easy to obtain converged results for
angle-integrated partial cross sections, special care is needed to obtain
converged results for differential cross sections at higher energies. 
Coverged results for partial cross sections are obtained for
$J_{\mbox{max}}=30$ at all energies. For obtaining convergent differential
cross sections, we need to take $J_{\mbox{max}}=150$ partial waves at 100
eV. However,
$J_{\mbox{max}}=30$ is sufficient for obtaining convergent differential
cross sections at 20 and 30 eV.

Here we present results of Ps-He scattering using the three-Ps-state model 
that includes the 
 following states: Ps(1s)He(1s1s), Ps(2s)He(1s1s), and 
Ps(2p)He(1s1s).  
 The Born terms
for the   excitation of  He  are found to be 
 small and are not
considered here in the coupled-channel scheme.  
First, we  present the elastic
Ps(1s)He(1s1s)  differential cross section and inelastic differential
cross
sections to
Ps(2s)He(1s1s) and  Ps(2p)He(1s1s) states at different energies.

In order to show the general trend of the differential cross sections, we
perform calculations at the following incident positronium energies: 20,
30, 40, 60, 80 and 100 eV.  We exhibit the differential cross sections for
elastic scattering at these energies in Fig. 1.  In Figs. 2 $-$ 3 we show
the inelastic cross sections for transition to Ps(2s)He(1s1s) and
Ps(2p)He(1s1s)  states. From all these figures we find that, as expected,
the differential cross sections are more isotropic at low energies where
only the low partial waves contribute.  At higher energies more and more
partial waves are needed to achieve convergence and the differential cross
sections are more anisotropic. The small oscillation of the differential
cross sections at larger angles and energies is due to numerical
difficulties.  

Recently, Garner et al. \cite{3x} have provided an experimental estimate
of average differential cross section across the energy range 10 to 100 eV
with respect to any process in Ps-He scattering for forward scattering
angles: $\langle d\sigma/d\Omega \rangle =(34\pm 12)\times 10^{-20}$ m$^2$
sr$^{-1}$ =
$(121\pm 43) a_0^2$ sr$^{-1}$. However, it is not possible to make a
meaningful comparison between the present differential cross sections and
the experimental estimate  of Garner et al.

We  calculate the different angle-integrated partial cross sections
for Ps-He scattering. In addition to the Ps(1s,2s,2p) cross sections
calculated using the coupled-channel method, we also calculate the higher
Ps($7>n>2$)-excitation and Ps-ionization cross sections using the Born
approximation with present exchange potential. These results are shown in
Fig. 4, where we plot angle-integrated elastic, Ps($n$=2)
[$\equiv$Ps(2s+2p)], inelastic
Ps($7>n>2$), and Ps ionization cross sections. The total cross section
calculated from these partial cross sections is also shown in this plot
and compared with the experiments of Refs. \cite{1,2} and total cross
section of the 22-Ps-state R-matrix calculation of Ref. \cite{4a}. The
agreement between theory and experiment is quite good up to 70 eV. The
target-inelastic processes ignored in this work are supposed to play
important role at higher energies, which may be  the cause of
detorioration of agreement of present results with experiment
above 70 eV. There exists qualitative disagreement between the present
total cross section and that of the 22-state calculation of Ref.
\cite{4a}, on which we comment below.

As the Ps-He system is of fundamental interest to both theoreticians and
experimentalists, it is appropriate to critically compare our results with
other theories and experiments.  The only other recent experiment on Ps-He
is the one by Nagashima et al.  \cite{3}, who obtained the cross section
of ($13\pm 4)\pi a_0^2$ for an average energy of 0.15 eV in striking
disagreement with the present calculation yielding $2.58\pi a_0^2$ at 0.9
eV as well as with the experiment of Skalsey et al. \cite{2} who obtained
$(2.61\pm 0.5)\pi a_0^2$ at about 0.9 eV.

Independent experiment on the
measurement\cite{22} of pick-off
quenching rate of Ps on He can be used \cite{23} to resolve the
stalemate.
It is argued \cite{23} that a large low-energy Ps-He elastic cross
section
implies a large repulsive exchange potential between Ps and He atoms in
the elastic channel.  In the presence of a  large repulsive
potential it will be difficult for the  Ps atom to approach the 
 He atom. Consequently, one will have 
a small value for the pick-off quenching rate. From a
study
of the pick-off quenching rates of different models, we concluded
\cite{23} 
 that a small low-energy cross section, as  obtained by us,   will lead to
a
large pick-off quenching rate in agreement with experiment. The large
low-energy cross sections as obtained in other theoretical models
\cite{4,4a,5x,5y} will lead
to a much too small pick-off quenching rate in disagreement with
experiment.
This substantiates  that the present low-energy cross section and the
experiment of Skalsey et al. \cite{2} are consistent with the pick-off
quenching rate 
measurement \cite{22}. It would be difficult to reconcile the low-energy
cross
section of
Nagashima et al. \cite{3} and other theoretical results \cite{4,4a,5x,5y}
with the measurement of the 
pick-off
quenching 
rate. 

We note that a model calculation by G. Peach \cite{24},
performed
before the experiment of Skalsey et al. \cite{2},   is also in
reasonable agreement with the present calculation and low energy
experiments. The model of Peach was constructed by fitting to known
positron-helium \cite{25} and electron-helium \cite{26} scattering data.
 
	In Table I we compare the results of the 
angle-integrated partial cross sections to Ps(1s,2s,2p) states  of
different theoretical
calculations. The present Ps(1s) Born cross sections are much smaller than
the
Born-Oppenheimer cross sections \cite{19} used as input to close-coupling
\cite{4} or
R-matrix \cite{4a} schemes. There have been different static-exchange
calculations on Ps-He 
since the 1960s \cite{4,4a,5x,5y}. These calculations yielded similar
results
and in the
static-exchange (SE) column of Table I we quote the recent cross sections
of Refs.
\cite{4,4a}. Although these SE cross sections are much smaller than the
corresponding Born-Oppenheimer cross sections, they are much larger than
those of the present calculation. The 22-Ps-state R-matrix
calculation \cite{4a} yields elastic cross sections marginally smaller
than the SE
cross
sections, and it seems unlikely that the ``converged" R-matrix
calculation will lead to elastic cross sections comparable to the present
ones. 
However, the measured pick-off quenching rate \cite{22} favors \cite{23}
a week exchange
potential and small Ps(1s) cross sections at low energies, and  future
measurements of
low-energy Ps-He elastic cross sections will decide which of the results
are more
realistic.
Although the present elastic Ps(1) cross sections are much smaller than
those of the R-matrix calculation, the reverse is true for the excitation
cross sections to the Ps(2) states as can be found from Table I. The
large Ps-excitation (and Ps ionization) cross sections of the present
calculation  and the small low-energy elastic cross sections are
collectively 
responsible  for the construction of the pronounced peak in the total
cross section as in Fig. 4 near  15 $-$ 20 eV in agreement with
experiments of Refs.
\cite{1} and \cite{2}. This  peak is also present in the
calculation of Peach \cite{24} and 
 is clearly absent in the close-coupling \cite{4} and 22-Ps-state R-matrix
analysis \cite{4a}. Similar  peaks also appear in the total cross
section of Ps-H$_2$ and Ps-Ar scattering \cite{3x}.

To summarize, we have performed a three-Ps-state coupled-channel
calculation of Ps-He scattering at low and medium energies using a
regularized symmetric nonlocal electron-exchange model potential recently
suggested
by us and successfully used in other Ps scattering problems. We present
results for differential cross sections at several incident Ps energies
between 20 eV to 100 eV for elastic scattering and inelastic excitation to
Ps(2s,2p)He(1s1s) states. We also present the angle-integrated partial
cross sections and compare them with those of other calculations. The
present total cross sections are in agreement with data of Refs.
\cite{1,2}. However, there is alarming discrepancy between the present
cross sections and those of conventional R-matrix \cite{4a} and
close-coupling \cite{4} calculations. These latter calculations are in
agreement with a recent measurement of low-energy cross section by
Nagashima et al. \cite{3}. 
At low energies, the present elastic cross
sections are much too smaller compared to those of Refs. \cite{4,4a}. 
However, the present total cross section develops a pronounced maximum
near 15 $-$ 20 eV as can be seen in Fig. 4 in agreement with the general
experimental trend \cite{3x}. The cross section of Ref. \cite{4a} does not
have this
behavior.  Although, comparison with the pick-off quenching measurement
data \cite{22} at low-energy favors \cite{23} the results of the present
model, further precise measurements of total and Ps(2) excitations at low
energies will finally resolve the stalemate.

The work is supported in part by the Conselho Nacional de Desenvolvimento -
Cient\'\i fico e Tecnol\'ogico,  Funda\c c\~ao de Amparo
\`a Pesquisa do Estado de S\~ao Paulo,  and Finan\-ciadora de Estu\-dos e
Projetos of Brazil.

\newpage
\vskip .2cm
{Table I: Angle-integrated Ps-He partial cross sections in $\pi a_0^2$ at
different positronium energies:  EB $-$ first Born with present
exchange; BO $-$ first Born with Born-Oppenheimer 
exchange;
SE $-$ static exchange of Refs.
\cite{4,4a};
3St $-$ three-Ps-state with
present exchange; 22St  $-$ 22-Ps-state R-matrix calculation of Ref.
\cite{4a}}

\vskip .2cm

\begin{centering}

\begin{tabular} {|c|c|c|c|c|c|c|c|c|c|c|c|} \hline Energy  & Ps(1s) &
Ps(2s)
&
Ps(2p)&Ps(1s) & Ps(1s)&Ps(1s) &
 Ps(1s) & Ps(2s)  & Ps(2p) & Ps(2) & Ps(2) \\(eV) & EB & EB &
EB&BO&
SE& 22St  & 3St & 3St & 3St &
3St  & 22St\\
  \hline
 0 & 15.82 & & &&14.6 & 13.2 & 3.34   & & & &\\
 0.068 & 15.33 & &  &132 &14.4 & 13.0 & 3.15&  & & & \\
 0.612 & 12.11& & &98&  12.9& &2.75 &  & &  &  \\
 1.088 & 10.04& & &78& 12.1&11.3&2.48 &  &  & &  \\
1.7 & 8.08& & &59& 11.3& &2.18 & & & &  \\
2.448 &6.38& & &44&  10.5&9.4  &  1.88 & & & &  \\
4.352 & 3.91& & &23& 9.0& &1.26 & & & & \\
5 & 3.39&  & &  &8.6 &7.1 &1.00&      & & & \\
5.508 &3.06 &0.070 &1.44&  &
& &0.96 &0.071 &1.15 &1.22  & 0.24\\
6 & 2.79& 0.091&1.78 & & 8.1& 6.1 &0.97         &0.083 &1.35
&1.43
&0.42 
\\
6.8 & 2.42& 0.100& 1.89& 12&7.7&  &0.96 &0.074 &1.47 & 1.54 &
\\
8 &1.99&0.097&1.77& &7.1 & 4.8 &0.92 &0.056 & 1.45&  1.51
&0.50
\\
10 & 1.51 & 0.080 & 1.48 & &6.7 & 3.8 &0.84 &0.048 & 1.29 &1.34
&0.51\\
15 &0.86&0.048&0.97&3.0 & 4.8 & 2.4 &0.63 & 0.042 &  0.91& 0.95
&
0.44
\\
20 & 0.56&0.031&0.70&1.7 &3.6&1.5 &0.46 &0.032 &0.67 & 0.70 &
0.31
\\
30 &0.29&0.016&0.43& 0.6&2.0& 1.0 &0.27&0.017 &0.42 &  0.44
&0.18
\\
40 &0.17&0.0094&0.30&0.22 &0.7&0.8 &0.17 &0.010 &0.30 & 0.31
&0.12  
\\
50 & 0.11&0.0061&0.23& & & &0.11 &0.0067 &0.23 & 0.24 & 
\\60 &0.079&0.0042&0.18&0.04& 0.08& &0.077
&0.0045&
0.18 &0.19 & \\
80 &0.043& 0.0023&0.13& 0.007&0.01&  &
0.042 &0.0024  &
0.13
&0.13  &  \\
100 &0.026&0.0014&0.10&0.001&0.002&
 &0.026 &0.0014&0.10
&0.10
 &  \\
150 &0.010&0.0005&0.06&  & &
 &0.010 &0.0005&0.06
&0.06
 &  \\
\hline
\end{tabular}

\end{centering}

\newpage
{\bf Figure Caption:}

1. Differential cross section (in units of $a_0^2$) for elastic Ps-He
scattering at the following incident Ps energies: 20 eV (dashed-dotted
line), 30 eV (dashed-double-dotted line), 40 eV (dashed-triple-dotted
line), 60 eV (full line), 80 (long dashed line), and 100 eV (short dashed
line).

2. Differential cross section (in units of $a_0^2$) for inelastic
Ps-He
scattering to Ps(2s)He(1s1s) state at the following incident Ps energies:
20
eV (dashed-dotted
line), 30 eV (dashed-double-dotted line), 40 eV (dashed-triple-dotted
line), 60 eV (full line), 80 (long dashed line), and 100 eV (short dashed
line). 

3. Differential cross section (in units of $a_0^2$) for inelastic
Ps-He
scattering to Ps(2p)He(1s1s) state at the following incident Ps energies:
20
eV (dashed-dotted
line), 30 eV (dashed-double-dotted line), 40 eV (dashed-triple-dotted
line), 60 eV (full line), 80 (long dashed line),  and 100 eV (short dashed
line). 

4. Partial and total cross sections (in units of $10^{-16}$ cm$^2$)
of Ps-He scattering at different Ps energies: Ps(1s)
(dashed-triple-dotted line), Ps($n$=2) (dashed-dotted line), Ps($7>n>2$)
(dashed-double-dotted line), Ps-ionization (dashed line), total (full
line), total (full line with crosses from Ref. \cite{4a}), and 
data points with error bars from Refs. \cite{1,2}.

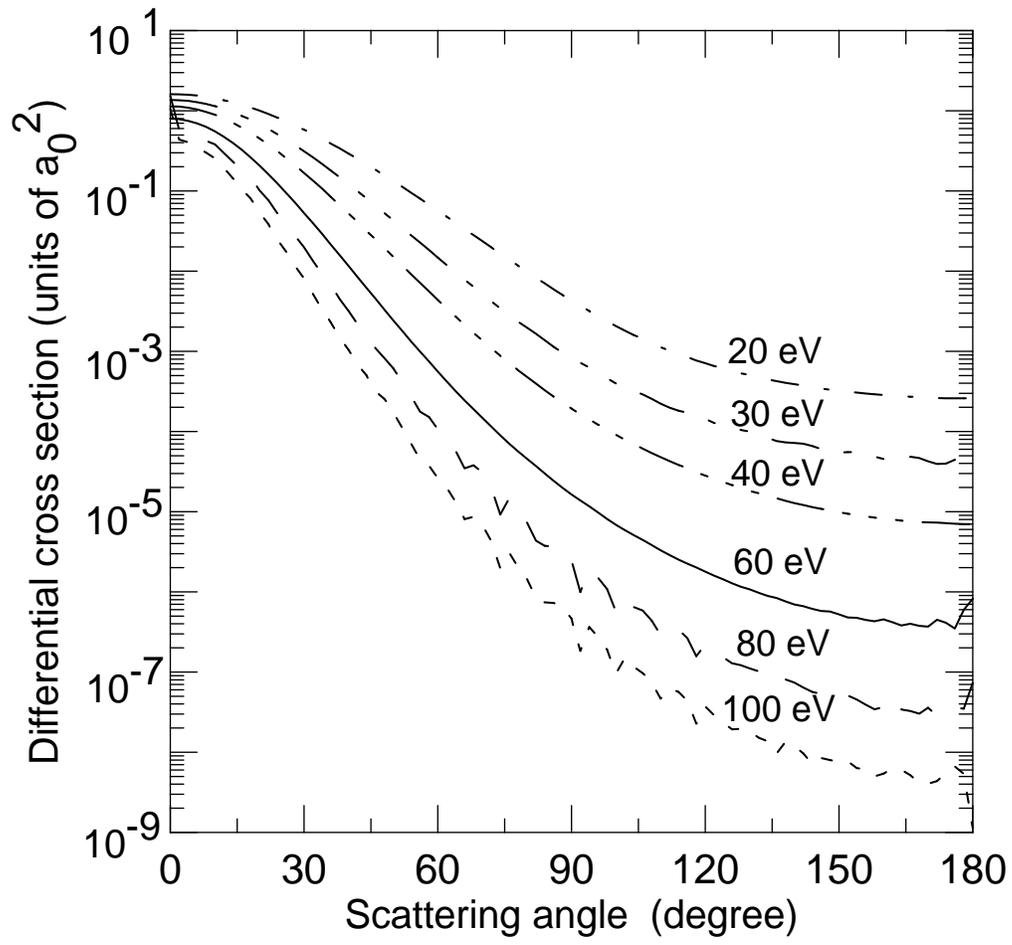
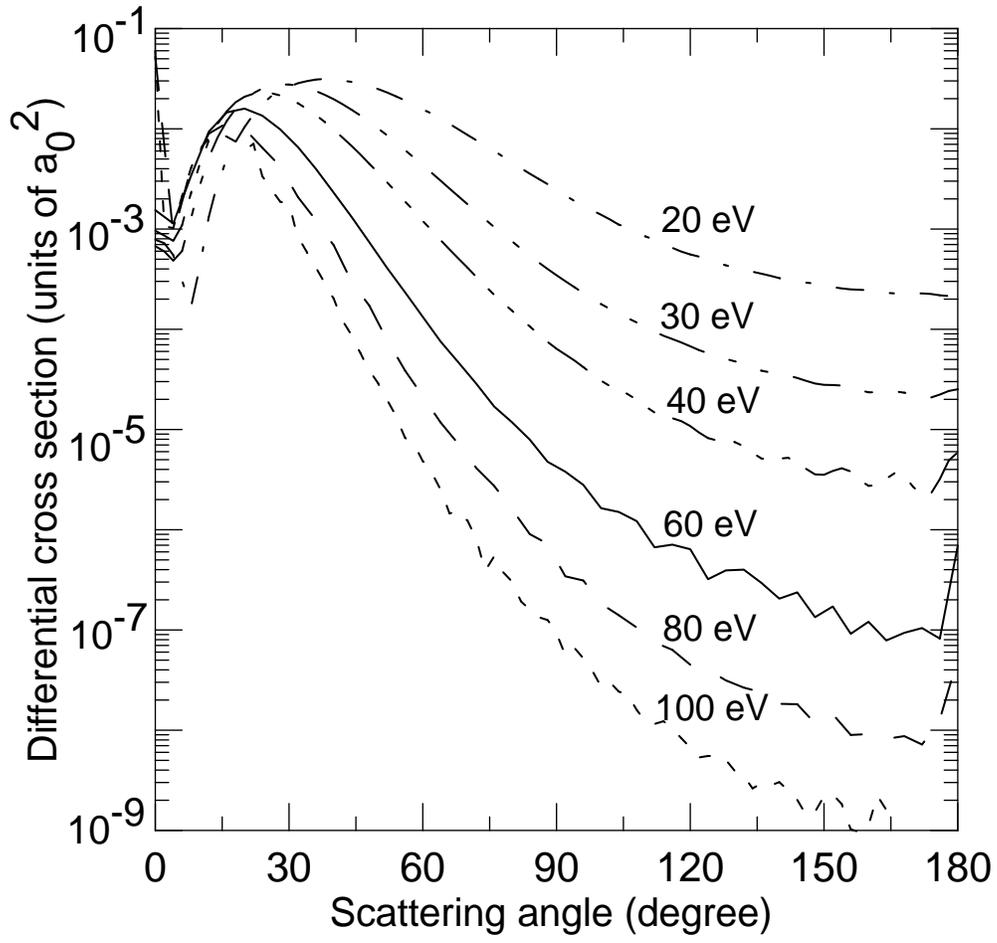
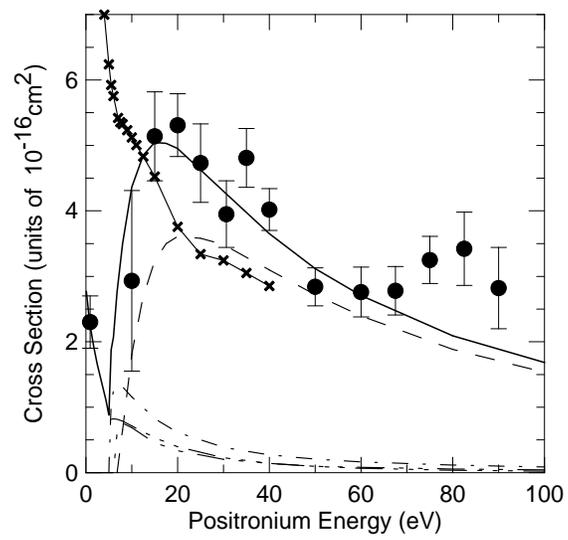
\end{document}